\documentstyle[12pt]{article}

\begin{document}

\begin{center}
\bf {\Large $^1S_0$ Proton and Neutron Superfluidity in
$\beta$-stable Neutron Star Matter}
\end{center}

 \begin{center}
 W. Zuo$^{a,b,}$\footnote{Corresponding address:
Institute of Modern Physics, Chinese Academy of Science, P.O.Box
31, Lanzhou 730000, China. Tel: 0086-931-4969318; E-mail:
zuowei@impcas.ac.cn},
Z. H. Li$^{a}$, G. C. Lu$^{a}$, J.Q.Li$^{a,b}$, W. Scheid$^{b}$\\
 U. Lombardo$^{c,d}$,H.-J. Schulze$^{e}$, and C. W. Shen$^{d}$\\[5mm]

{\small $^{a}$ Institute of Modern Physics, Chinese Academy of
Science,
Lanzhou 730000, China \\
$^{b}$ Institute f\"ur Theoretische Physik der
Justus-Liebig-Universit\"at, D-35392 Giessen, Germany\\
$^{c}$ Dipartimento di Fisica, Universit\`a di Catania,
       57 Corso Italia, I-95129 Catania, Italy \\
$^{d}$ Laboratori Nazionali del Sud,
       Via Santa Sofia 44, I-95123 Catania, Italy \\
$^{e}$ INFN, Sezione di Catania,
       57 Corso Italia, I-95129 Catania, Italy }
\end{center}

\begin{abstract}
We investigate the effect of a microscopic three-body force on the
proton and neutron superfluidity in the $^1S_0$ channel in
$\beta$-stable neutron star matter. It is found that the
three-body force has only a small effect on the neutron $^1S_0$
pairing gap, but it suppresses strongly the proton $^1S_0$
superfluidity in $\beta$-stable neutron star matter.\\[2mm]
{\bf PACS numbers:} 26.60.+c,21.65.+f,
      21.30.Fe \\[2mm]
{\bf Keywords:} $^1S_0$ superfluidity, three-body Force,
$\beta$-stable neutron star matter, Brueckner-Hartree-Fock
approach
\end{abstract}


Superfluidity plays an important role in understanding a number of
astrophysical phenomena in neutron
stars~[1-10].
It is generally expected that the cooling processes via neutrino
emission~\cite{PAGE,HEIS,LINK}, the properties of rotating
dynamics, the post-glitch timing observations~\cite{SHAP,SAUL},
the possible vertex pinning~\cite{PINE} of a neutron star are
rather sensitive to the presence of neutron and proton superfluid
phases as well as to their pairing strength. For instance, since
the paired nucleons do not contribute to thermal excitations, the
proton superfluidity could suppress considerably the neutrino
emission processes, and consequently affects the neutrino cooling
rate of a neutron star remarkably.

Since the neutron and proton superfluidity properties in neutron
stars are related only indirectly to the observations, reliable
and precise theoretical predictions based on microscopic many-body
approaches are highly desirable. In Refs.~\cite{BALDO,ELGAR1} it
is reported by using the Brueckner-Hartree-Fock ( BHF )
calculations based on purely two-body nucleon-nucleon ( $NN$ )
interactions that the $^1S_0$ neutron superfluid can be formed
only in the low-density region ( $\rho_B<0.1fm^{-3}$ ) with a
maximal gap value of about
 $2.8$MeV at a Fermi momentum $k_F\simeq 0.8$fm$^{-1}$,
 while due to the small proton fraction in
$\beta$-stable matter the proton superfluid phase in the $^1S_0$
channel may extend to much higher baryon densities up to
$\rho_B\sim 0.4$fm$^{-3}$ with a maximal pairing gap of about
0.9MeV. In Ref.~\cite{ELGAR2}, the relativistic effects on the
superfluidity in $\beta$-stable neutron star matter have been
investigated by using the Dirac-BHF ( DBHF ) approach and the
one-boson-exchange $NN$ interaction ( BONN potential
)~\cite{BONN}. It was found that the relativistic effect on the
proton superfluidity in the $^1S_0$ channel is very weak, while
the magnitude of the neutron pairing gap in the $^3P_2$ channel is
reduced strongly due to relativistic corrections to the
single-particle ( s.p.~) energies.

The pairing correlations in nuclear medium are essentially related
to the underlying $NN$ interaction and their magnitude is
determined by the competition between the repulsive short-range
and attractive long-range parts of the interaction. The solution
of the gap equation is extremely sensitive to the medium
modifications of both the bare $NN$ interaction and the s.p.
 energy. These medium effects on nuclear pairing are extremely
difficult to study on a microscopic level and have motivated an
extensive investigation by many authors~\cite{MEDI1,SELF}.
Three-body forces, which turn out to be crucial for reproducing
the empirical saturation properties of nuclear matter in a
non-relativistic microscopic approach~\cite{math,lej,mac}, are
expected to modify strongly the in-vacuum $NN$ interaction,
especially the short-range part~\cite{math,mac,ZUO}. However,
their effects on the superfluidity properties in neutron stars
have not yet been well investigated. The aim of this letter is
devoted to the influence of three-body forces on the $^1S_0$
neutron and proton superfluid phases in $\beta$-stable matter.

For such a purpose, we shall not go beyond the BCS framework. In
this case, the pairing gap which characterizes the superfluidity
in a homogeneous Fermi system is determined by the standard BCS
gap equation~\cite{RING}, i.e.,
\begin{eqnarray}\label{e:gap}
 \Delta_{\vec{k}} = -\sum_{\vec{k'}}
 v(\vec{k},\vec{k}') \frac{1}{2E_{\vec{k}'}}
 {\Delta_{\vec{k}'}} \:,
\end{eqnarray}
where $v(\vec{k},\vec{k}')$ is the bare $NN$ interaction in
momentum space,
$E_{\vec{k}}=\sqrt{(\epsilon_{\vec{k}}
-\epsilon_F)^2+\Delta_{\vec{k}}^2}$, $\epsilon_{\vec{k}}$ and
$\epsilon_F$ being the s.p. energy and its value at the Fermi
surface, respectively.

In the BCS gap equation, the most important ingredients are the
$NN$ interaction $v(\vec{k},\vec{k}')$ and the neutron and proton
 s.p. energies $\epsilon_{\vec{k}}$ in $\beta$-stable matter. For
the $NN$ interaction, we adopt the Argonne $AV18$ two-body
interaction~\cite{wir} plus a microscopic three-body force
 (~TBF~)~\cite{math}. The TBF is constructed self-consistently
  with the
$AV18$ two-body force by using the meson-exchange current
approach~\cite{lej} and it contents the contributions from
different intermediate virtual processes such as virtual
nucleon-antinucleon pair excitations, and nucleon resonances ( for
details, see Ref.~\cite{math}). The TBF effects on the equation of
state ( EOS ) of nuclear matter and its connection to the
relativistic effects in the DBHF approach have been reported in
Ref.~\cite{lej}. The influence of the TBF on Landau parameters in
nuclear matter and on the neutrino mean free path in neutron stars
has also been explored~\cite{ZUO03}.

The proton fraction and the s.p. energies in $\beta$-stable matter
are calculated by using the BHF approach for isospin asymmetric
nuclear matter~\cite{ZUO1}. In solving the Bethe-Goldstone
equation for the $G$-matrix, the continuous choice~\cite{MAHA0}
for the auxiliary potential is adopted since it provides a much
faster convergence of the hole-line expansion than the gap
choice~\cite{SONG}.
The effect of
the TBF is included in the self-consistent Brueckner procedure
along the same lines as in Refs.~\cite{math,lej}, where an
effective two-body interaction is constructed to avoid the
difficulty of solving the full three-body problem. A detailed
description and justification of the method are discussed in
Refs.~\cite{math,lej}. Here we simply write down the equivalent
two-body potential in $r$-space
\begin{eqnarray}\label{e:tbf}
& & \langle\vec r_1 \vec r_2| V_3 |
  \vec r_1^{\ \prime} \vec r_2^{\ \prime} \rangle =
  \frac{1}{4} Tr\sum_n \int {\rm d}{\vec r_3}
  {\rm d} {\vec r_3^{\ \prime}}
\phi^*_n(\vec r_3^{\ \prime})\phi_n(r_3) (1-\eta(r_{23}'))
(1-\eta(r_{13}' )) \nonumber
\end{eqnarray}
\begin{eqnarray}
\times W_3(\vec r_1^{\ \prime} \vec r_2^{\ \prime} \vec r_3^{\
\prime}|\vec r_1 \vec r_2 \vec r_3)
 (1-\eta(r_{13}))(1-\eta(r_{23}))\ ,
\end{eqnarray}
where the trace is taken with respect to the spin and isospin of the
third nucleon, and $\eta(r)$ is the defect function. According to
Eq.~(\ref{e:tbf}) the effective two-body force is obtained by
averaging the three-body force over the wave function of the third
nucleon taking into account the correlations between this nucleon
and the two others. Due to its dependence on the defect function
the effective two-body force is calculated self-consistently along
with the $G$-matrix and the auxiliary potential at each step of
 the iterative BHF procedure.

 With the obtained EOS of asymmetric nuclear matter the proton
fraction $Y^p$ for a given total baryon density $\rho_B$ in
$\beta$-stable matter can be calculated according to the
charge-neutrality and the equilibrium condition with respect to
weak interaction~\cite{lej}.
 The calculated proton fractions are listed
 in Tab.~1, where the first column gives the total baryon
 densities, the second column
and the third column present the corresponding proton fractions
obtained by using the $AV18$ two-body interaction and the $AV18$
plus the TBF, respectively. Inclusion of the TBF in the
calculations increases the proton fractions at high densities.
This has to be attributed to the influence of the TBF on the EOS
of asymmetric nuclear matter as verified in Ref.~\cite{lej}, where
the possible implications for the neutron star cooling mechanisms
were also discussed.

To solve the gap equation, we follow the scheme given in
Ref.~\cite{BALDO}, where it is shown that the gap equation can be
split into two coupled equations,
\begin{eqnarray} \label{e:gap2}
 \Delta_{\vec{k}} & = & -\sum_{k'\le k_c}
 \widetilde{V}(\vec{k},\vec{k'})
{1 \over 2E_{\vec{k'}} } \Delta_{\vec{k}'}\:, \\
 \widetilde{V}(\vec{k},\vec{k'}) &=&  V(\vec{k},\vec{k'})
 - \sum_{k'' \geq k_c}
{  V(\vec{k},\vec{k''}) \over 2E_{\vec{k''}} }
  \widetilde{V}(\vec{k''},\vec{k'})
  \:,
\end{eqnarray}
where the effective interaction $\widetilde{V}$, arising from the
introduction of a cutoff $k_c$ in momentum space, sums up a series
of ladder diagrams analogous to the Bethe-Goldstone equation and
it is quite sensitive to the tail ($k > k_c$) of the $NN$
interaction, which reflects the short-range part of the nuclear
force.

\begin{center}
Table 1 \ Proton fractions $Y^p$ in $\beta$-stable matter. \\[4mm]
\begin{tabular}{ c c c }
\hline\hline $\rho_B (fm^{-3})$ & \multicolumn{1}{c}
~~~~~~~~~~~~~~~~~~~~
$Y^p$ \\
 \hline\hline    &
BHF (AV18) & BHF (AV18 + TBF)
\\ \hline
 0.005 &0.0042           & 0.0042             \\
 0.009 &0.0078           & 0.0077             \\
 0.020 &0.0093           & 0.0091             \\
 0.030 &0.0138           & 0.0132             \\
 0.050 &0.0187           & 0.0181             \\
 0.070 &0.0225           & 0.0218             \\
 0.085 &0.0247           & 0.0252             \\
 0.100 &0.0279           & 0.0280             \\
 0.140 &0.0332           & 0.0353             \\
 0.170 &0.0382           & 0.0432             \\
 0.210 &0.0471           & 0.0570             \\
 0.250 &0.0558           & 0.0731             \\
 0.300 &0.0667           & 0.0944             \\
 0.340 &0.0746           & 0.1162             \\
 0.400 &0.0872           & 0.1499             \\
 0.450 &0.0970           & 0.1785             \\
\hline \hline
\end{tabular}
\end{center}

In order to numerically investigate the effect of the TBF we have
solved the gap equation,
 by adding the effective TBF given in Eq.(\ref{e:tbf})
 to the bare $AV{18}$
two-body force. At the same time the
 s.p. energy spectrum $\epsilon_{\vec{k}}$ appearing in the gap
equation is computed from the BHF approach by using the $AV18$
plus the same TBF.

 Fig.~\ref{f:neutrongap} shows the neutron energy gap in
the $^1S_0$ partial wave channel $\Delta_F=\Delta(k_F)$ as a
function of the total baryon density $\rho_B$. The dashed curve is
obtained by adopting the pure $AV18$ two-body
interaction only, while the solid curve is predicted by using the
$AV18$ plus the TBF. For comparison, the dotted and dot-dashed curves
 are the energy gaps reported in Ref.~\cite{BALDO} using
the Argonne $AV14$ and the Paris potentials, respectively.
It is seen that the results calculated with the three
different two-body interactions agree well with each other, i.e.,
the neutron superfluidity phase in the $^1S_0$ channel can only
occur in the low-density region ($\rho_B< 0.1$fm$^{-3}$) of neutron
stars with a maximal gap value of about 2.8MeV peaked at a Fermi
momentum $k_F\simeq 0.8$fm$^{-1}$ (the corresponding total baryon
density is $\rho_B\simeq 0.02$fm$^{-3}$). The TBF effect is quiet
small, i.e.,
 almost negligible at relatively low density and a slight suppression
of the gap as increasing density. This result is expected from the
low density for the $^1S_0$ neutron superfluidity, since
three-body forces are invented to take, in an effective way, the
non-nucleonic degrees of freedom in nuclear medium into account
and become significant only at high densities, i.e., around and
above the empirical saturation density~\cite{mac}.

In Fig.~\ref{f:protongap} is reported the proton $^1S_0$ energy
gap in $\beta$-stable matter. The solid curve is obtained by
including the TBF while the dashed one by using the pure $AV18$
two-body force only. Without the TBF, our result is in good
agreement with the predictions by adopting the BONN
potential~\cite{ELGAR1,ELGAR2}, the $AV14$ and the Paris
potentials~\cite{BALDO}. As compared to the neutron $^1S_0$
superfluidity, the proton $^1S_0$ superfluid phase extends to much
higher densities but with a smaller peak gap value around
$\rho_B\sim 0.2$fm$^{-3}$. The former is a direct consequence of
the small proton fraction in $\beta$-stable matter, and the latter
stems from the different s.p. potentials for neutrons and for
protons~\cite{BALDO}. As is known, in isospin highly asymmetric
nuclear matter like $\beta$-stable matter, the proton s.p.
 potential is much deeper than the neutron one~\cite{lej,ZUO}.

The effects of the TBF are twofold as shown in
Fig.~\ref{f:protongap}. One is a strong reduction of
the peak value of the gap from $\sim 0.95$MeV to $\sim 0.55$MeV
and a remarkable shift of the peak to a much lower baryon density
from $\sim 0.2$ fm$^{-3}$ to $\sim 0.09$fm$^{-3}$. The another is
that the TBF leads to a noticeable shrinking of the density region of
the superfluid phase from $\rho_B\le 0.45$fm$^{-3}$ to
$\rho_B\le 0.3$fm$^{-3}$. The above predicted TBF suppression of
the $^1S_0$ proton superfluidity in $\beta$-stable matter appears
inconsistent with the small proton fractions which correspond to
small proton densities in the matter. However, since proton pairs are
embedded in the medium of neutrons and protons, both the
surrounding protons and neutrons contribute to the TBF
renormalization of the proton-proton interaction.
 This means that
the relevant density to the TBF effect
is the total baryon density, but not the
proton one. One can verify from Fig.~2 that the strongest
suppression of the energy gap is mainly in the region $\rho_B\ge
\rho_0$, $\rho_0$ being the empirical saturation density
of nuclear matter, and the reduction of the gap increases rapidly
as increasing the total baryon density.

It is worth noticing the discrepancy between the TBF effect and
the relativistic effect~\cite{ELGAR2} on the $^1S_0$ proton
superfluidity in $\beta$-stable matter. The investigation of
Ref.~\cite{ELGAR2} shows that the relativistic effect reduces
remarkably the $^3P_2$ neutron gap, but the differences between
the relativistic and nonrelativistic proton gaps in the $^1S_0$
channel are quite small.
 One possible reason concerns the more or less
different mechanisms involved. In the DBHF
approach~\cite{mac,DBHF2}, the medium renormalization of the bare
$NN$ interaction are taken into account via the Dirac spinor which
is dressed in nuclear medium (which can be traced to the virtual
excitation of nucleon-antinucleon pairs~\cite{BROWN}), but the
most important effect on the pairing gaps ( for instance, the
$^3P_2$ neutron gap ) comes from the relativistic modification of
the s.p. energies~\cite{ELGAR2}.

The TBF is expected to influence the superfluidity phases in
$\beta$-stable matter via three different ways. First it
renormalizes the bare nucleon-nucleon interaction. Second it
modifies the s.p. energy spectrum in the gap equation and finally
the inclusion of the TBF changes the predicted proton fractions as
shown in Tab.~1. To see which is the most important mechanism
responsible for the strong suppression of the $^1S_0$ proton
energy gaps, we include the TBF in the BHF calculations to obtain
the proton fractions and s.p. energies, but adopt only the pure
$AV18$ as the $NN$ interaction $v(\vec{k},\vec{k}')$ in the gap
equation, i.e., Eq.(1). The results are shown by the dotted curve
in Fig.2. One can see that the combined effects via the proton
fractions and the s.p. energies are relatively small, i.e., a
slight reduction, and mainly in the higher density region
($\rho_B>0.3$fm$^{-3}$) where the TBF modifications of the proton
fractions ( Tab.1 ) and the proton s.p. energies~\cite{lej} become
appreciably larger.
 Hence, the strongest effect stems from the TBF
renormalization of the $NN$ interaction in the medium.

In summary, we have investigated the influence of the TBF
 on the neutron and proton pairing gaps in the $^1S_0$
channel in $\beta$-stable neutron star matter. It is shown that
the TBF has only a weak effect on the neutron $^1S_0$
superfluidity phase, i.e., a slight reduction of the energy gap,
due to the low-density region concerned. However it suppresses
strongly the proton superfluidity in the $^1S_0$ channel induced
by the two-body $NN$ interaction. The peak value of the proton
$^1S_0$ energy gap is reduced by about 50\% from $\sim$ 0.95MeV to
$\sim$ 0.55MeV and shifted to a much lower density by the
inclusion of the TBF. The density region for the superfluid phase
is also remarkably shrunken as compared to the pure two-body force
prediction. It is shown that this suppression is mainly related to
the TBF renormalization of the $^1S_0$ two-body interaction.

Besides the TBF effects, the medium renormalizations of the $NN$
interaction and the s.p. energies, i.e., the screening effects
such as the polarization effects and dispersive effects, may also
influence largely the superfluidity properties in nuclear medium.
Up to now, all investigations in the
literature\cite{rev,LOMB,MEDI1,SELF} have predicted a reduction of
the BCS superfluidity gap in the $^1S_0$ channel. Therefore we
expect that the screening effects may further suppress the $^1S_0$
proton superfluidity in $\beta$-stable matter.

Since the relativistic effects lead to a strong suppression of the
$^3P_2$ neutron superfluidity in $\beta$-stable
matter\cite{ELGAR2}, the $^1S0$ proton superfluidity becomes
critical in determining the outcome of neutron star cooling. It is
expected in Ref.\cite{ELGAR2} that the main suppression (of the
neutrino production) comes from the superfluid proton in the
$^1S_0$ state. In this letter, the predicted suppression of the
$^1S_0$ proton pairing gap does not favor the proton superfluid
phase which is expected to suppress the modified URCA processes in
the interior of a neutron star. This is compatible with the recent
result of Link~\cite{LINK} derived from the observations of
long-period precession in isolated pulsars.


\section*{Acknowledgments}

We are indebted to Prof. J.~W.~Clark for drawing our attention to
the study of the TBF in relation with pairing. The work has been
supported in part by the Knowledge Innovation
Project(KJCX2-SW-N02) of CAS, the
NNSF (10235030, 10175082), the Major State Basic Research
Development Program (G2000077400), the Major Prophase Research
Project of Fundamental Research of the Ministry of Science and
Technology (2002CCB00200), of China and DFG, Germany.

\noindent
 {\bf Figure Captions:}
\begin{figure}[ht] \caption{ Neutron $^1S_0$
energy gap in $\beta$-stable matter.
} \label{f:neutrongap}
\end{figure}

\begin{figure}[ht] \caption{ Proton $^1S_0$ pairing gap in
$\beta$-stable matter. The solid curve is predicted by using the
$AV18$ plus the TBF, and the dashed curve by using the $AV18$
two-body force only. The dotted curve is calculated with the pure
$AV18$ for the $NN$ interaction in the gap equation and with
inclusion of the TBF in the BHF calculation. } \label{f:protongap}
\end{figure}


\begin{thebibliography}{MEDI2}

\bibitem{rev} For a recent review, see D.~J.~Dean and M.~Hjorth-Jensen,
  Rev.~Mod.~Phys. {\bf 75}, 607 (2003) and references therein.

\bibitem{LOMB} U. Lombardo and H.-J. Schulze,
 Lecture Notes in
  Physics, vol.~578, Eds. D. Blaschke, N. K. Glendenning and
  A. Sedrakian (Springer-Verlag, Berlin, 2001), p.~30.

\bibitem{PRAK}M.~Prakash, Phys.~Rep. {\bf 242}, 191 (1994).

\bibitem{PETH}  C.~J.~Pethick, Rev.~Mod.~Phys. {\bf 64}, 1133 (1992);
  C.~J.~Pethick and D.~G.~ Ravenhall, Annu. Rev. Nucl. Part. Phys.
  {\bf 45}, 429 (1995).

\bibitem{PAGE}D.~Page, Astrophys.~J. {\bf 428}, 250 (1994).
\bibitem{HEIS}H.~Heiselberg and M.~Hjorth-Jensen, Phys.~Rep.
  {\bf 328}, 237 (2000).
\bibitem{LINK}B.~Link, Phys.~Rev.~Lett. {\bf 91}, 101101(2003).

\bibitem{SHAP} S.~L.~Shapiro and S.~A.~Teukosky, {\em Black Holes, White
  Dwarfs and Neutron Stars} (Wiley, New York, 1983).

\bibitem{SAUL} J.~A.~Sauls, {\em Timing Neutron Stars},
  eds. H.~Ogelman and E.~P.~J.~van den Henvel,
  (Dordrecht, Kluwer, 1989) p.457.

\bibitem{PINE}D.~Pines and M.~A.~Alpar, Nature, {\bf 316}, 27 (1985).


\bibitem{BALDO}M.~Baldo, J.~Cugnon, A.~Lejeune and U.~Lombardo,
   Nucl.~Phys. {\bf A515}, 409 (1990).
\bibitem{ELGAR1}\O.~Elgaroy, L.~Engvik, M.~Hjorth-Jensen and
   E.~Osnes, Nucl.~Phys. {\bf A604}, 466 (1996).
\bibitem{ELGAR2} \O.~Elgaroy, L.~Engvik, M.~Hjorth-Jensen and
   E.~Osnes, Phys.~Rev.~Lett. {\bf 77}, 1428 (1996).
\bibitem{BONN}R.~Machleidt, Adv.~Nucl.~Phys. {\bf 19}, 185 (1989).



\bibitem{MEDI1} J.~W.~Clark, C.~G.~Kallman, C.-H.~Yang
  and D. A. Chakkalakal, Phys.~Lett. {\bf B61}, 331 (1976);
T.~L.~Anisworth, J.~Wambach and D.~Pines,
  Phys.~Lett. {\bf B222}, 173 (1989);
 C.~J.~Pethick and D.~G.~Rawenhall,
  Ann.~N.~Y.~Acad.~Sci. {647}, 503 (1991);
J.~Wambach, T.~L.~Anisworth and D.~Pines,
  Nucl.~Phys. {\bf A555}, 128 (1993);
J.~M.~C.~Chen, J.~W.~Clark, R.~D.~Dave
  and V.~V.~Khodel, Nucl.~Phys. {\bf A555}, 59 (1993);
A.~Rabhi, R.~Bennaceur, G.~Chanfray and P.~Schuck,
 Phys.~Rev. {\bf C66}, 064315 (2002);
A.~Schwenk, B.~Friman and G.~E.~Brown,
 Nucl.~Phys. {\bf A713}, 191 (2003).



 \bibitem{SELF} U.~Lombardo,
 in {\it Nuclear Methods and the Nuclear Equation of State},
  Ed. M.~Baldo, (World Scientific, Singapore, 1999), p.~458;
  P.~Bozek, Phys.~Rev. {\bf C62}, 054316 (2000);
  U.~Lombardo, P.~Schuck and W.~Zuo,
 Phys.~Rev. {\bf C64}, 021301(R), (2001);
 M.~Baldo and A.~Grasso,
 Phys.~Lett. {\bf B485}, 155 (2000);
 M.~Baldo, U.~Lombardo, H.-J.~Schulze, and W.~Zuo,
 Phys.~Rev. {\bf C66}, 054304, (2002);
 C.~W.~Shen, U.~Lombardo, P.~Schuck, W.~Zuo
 and N.~Sandulescu, Phys.~Rev. {\bf C67}, 061303(R) (2003).




\bibitem{math} P.~Grang\'e, A.~Lejeune, M.~Martzolff
  and J.-F.~Mathiot, Phys.~Rev. {\bf C40}, 1040 (1989).

\bibitem{lej} A.~Lejeune, U.~Lombardo and W.~Zuo,
  Phys.~Lett. {\bf B477}, 45 (2000);
  W.~Zuo, A.~Lejeune, U.~Lombardo and J.-F.~Mathiot, Nucl.~Phys.
  {\bf A706}, 418 (2002).

\bibitem{mac} R.~Machleidt, Adv.~Nucl.~Phys. {\bf 19}, 189 (1989).

\bibitem{ZUO} W.~Zuo, U.~Lombardo, H.~J.~Schulze, and C.~W.~Shen,
  Phys.~Rev. {\bf C66}, 037303, (2002).

\bibitem{RING} P.~Ring and P.~Schuck, {\it The Nuclear
  Many Body Problem} (Springer-Verlag, New York, 1980).

\bibitem{wir} R.~B.~Wiringa, V.~G.~J.~Stoks, and R.~Schiavilla,
  Phys.~Rev. {\bf C51}, 38 (1995).

\bibitem{ZUO03} W.~Zuo, C.~W.~Shen and U.~Lombardo,
  Phys.~Rev. {\bf C67}, 037301 (2003); U.~Lombardo, C.~W.~Shen,
  N.~Van Giai and W.~Zuo, Nucl.~Phys. {\bf A722}, 532c (2003).

\bibitem{ZUO1} W.~Zuo, I.~Bombaci and U.~Lombardo, Phys.~Rev.
  {\bf C60}, 024605, (1999).
\bibitem{MAHA0} J.~P.~Jeukenne, A.~Lejeune and C.~Mahaux, Phys.~Rep.
  {\bf 25C}, 83 (1976).
\bibitem{SONG} H.~Q.~Song, M.~Baldo, G.~Giansiracusa and
U.~Lombardo, Phys.~Rev.~Lett. {\bf 81}, 1584 (1998).


 \bibitem{DBHF2} B. ter Haar and R.~Malfliet, Phys.~Rep. {\bf 149}, 208
 (1987);
M.~Hjorth-Jensen, T.~T.~S.~Kuo and E.~Osnes,
  Phys.~Rep. {\bf 261}, 125 (1995);
F. de Jong and H. Lenske, Phys.~Rev. {\bf C58},
     890 (1998);
C.~Fuchs, T.~Waindzoch, A.~Faessler and D.~S.~Kosov,
  Phys.~Rev. {\bf C58}, 2022 (1998).
\bibitem{BROWN}G.~E.~Brown, W.~Weise, G.~Baym and J.~Speth,
  Comm.~Nucl.~Part.~Phys. {\bf 17}, 39 (1987).

\end{thebibliography}
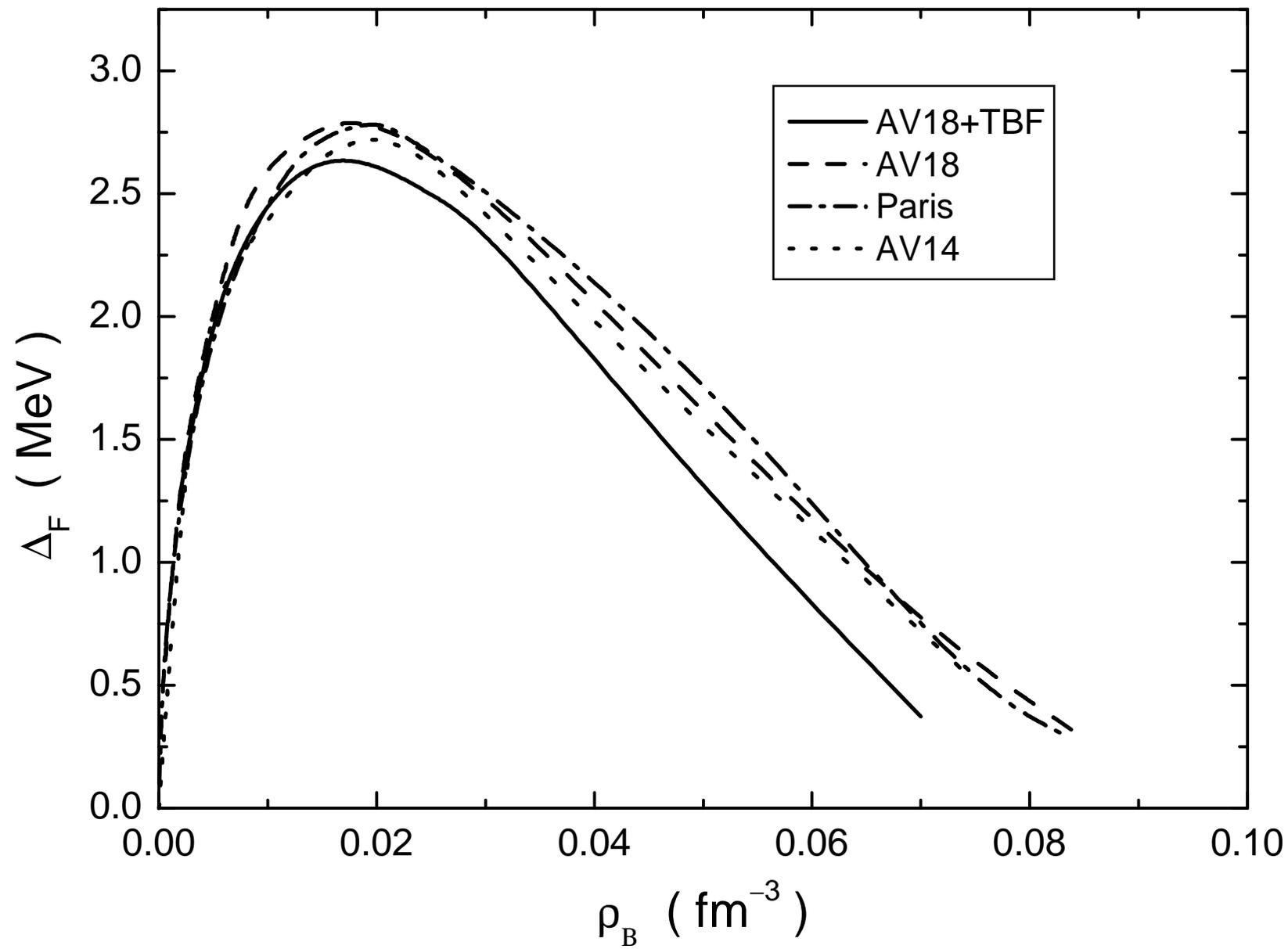
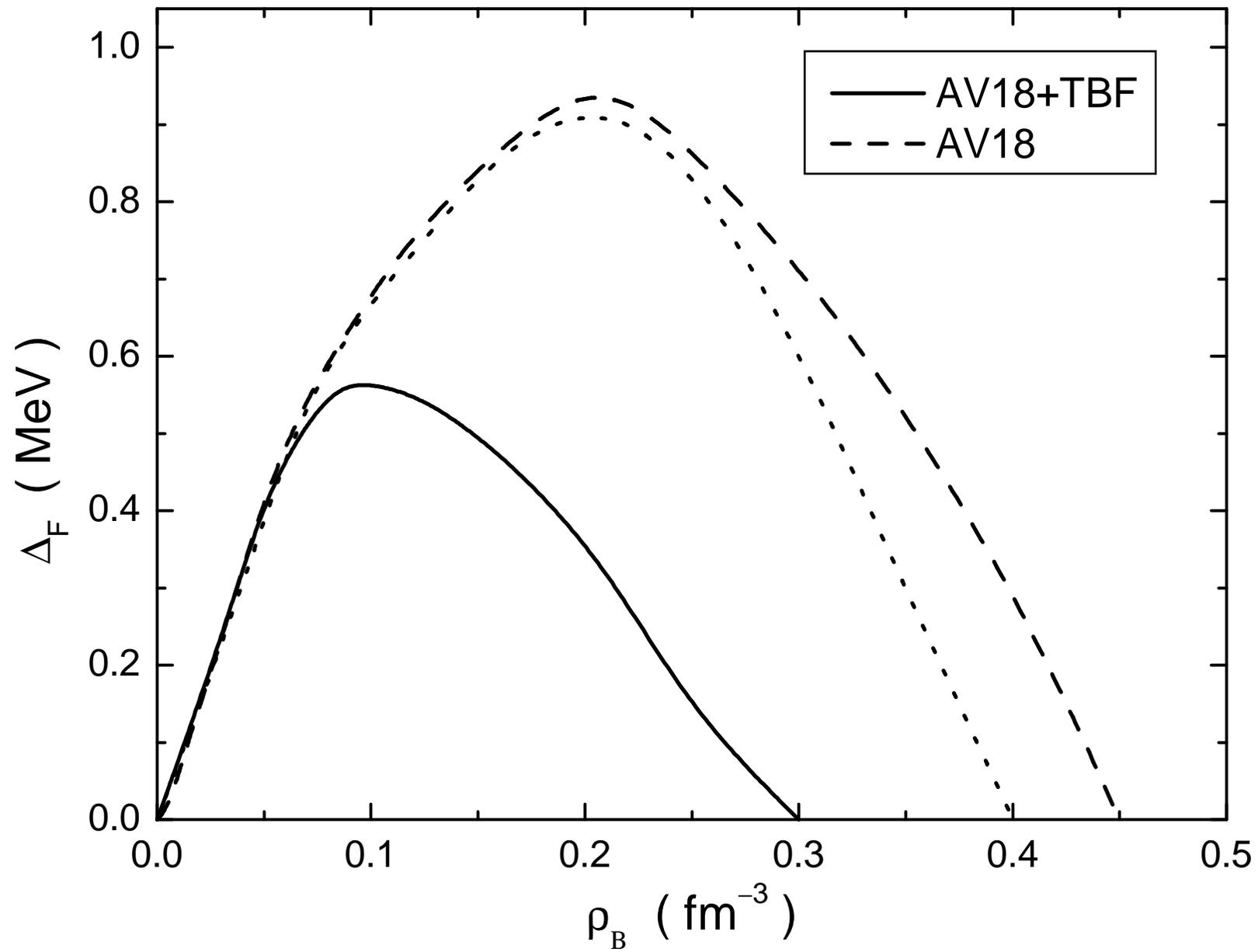
\end{document}